\documentclass[twocolumn,nofootinbib,aps,superscriptaddress,prd]{revtex4}
\usepackage{graphicx,dcolumn,bm,amsfonts,amsmath,color,xcolor,epsfig}
\usepackage{slashed}
\usepackage{graphics}
\usepackage{tikz}
\usepackage{multirow}
\usepackage{amsmath}
\usepackage{gensymb}
\usepackage{epstopdf}
\usepackage{hyperref}
\usepackage{amssymb}
\allowdisplaybreaks

\begin{document}

\title{Remarks on the $S$-wave masses of singly heavy mesons}

\author{Jisi Pan}
\email{ panjisi@gxstnu.edu.cn}
\address{School of Mechanical and Electrical Engineering, Guangxi Science $\&$ Technology Normal University,
Laibin 546199, China}
\author{Ji-Hai Pan\footnote{Corresponding author}}
\email{Tunmnwnu@outlook.com}
\address{College of Mathematics and Physics, Liuzhou Institute of Technology, Liuzhou 545000, China}

\begin{abstract}
Based on the study of the string model method of singly heavy mesons and singly heavy baryons, we calculate the mass spectrum of $1S$-and $2S$-wave for both charm and bottom mesons$(D/D_{s}, B/B_{s})$. Experimentally, there are most masses spectra of $1S$-wave have been found, while the masses part of the $2S$-state is not determined. In this paper, we will use singly light quark or diquark model images and Regge trajectory models, combined with perturbation processing methods, to analyze and study the observed singly heavy mesons, further predict the unobserved mesons masses and their corresponding spin-parity quantum numbers.
\end{abstract}

\maketitle

\section{introduction}

The $D/D_{s}$ mesons with charm quark and the $B/B_{s}$ mesons with bottom quark are typical heavy-light mesons, which is structurally analogous to a
hydrogen-like atoms( the singly light antiquark and the heavy quark resemble the extranuclear electron and the proton, respectively). In recent decades, more and more singly heavy (SH) mesons states $Q\bar q$ $(Q=c,b;\ \bar q=\bar n, \bar s)$ have been discovered by BaBar, Belleale, CLEO, and the LHCb experiment \cite{Tanabashi:PP88}, so the study of heavy-light mesons system has been attracting great attention. It can be seen from PDG \cite{Olive:PP88} that the experimental values of some low energy $D$, $D_{s}$ mesons \cite{Olive:PP888} have been basically determined. Since the experimental observation of the singly charm mesons states $D(2550)$, $D^{\ast}(2600)$, $D(2750)$and $D^{\ast}(2760)$, there have been different research methods \cite{Moir:PP88,Lewis:PP88,Godfrey:PP88,Godfrey:PP888,JiaD:PP888,JiaDH:PP888,Canonical:PP888} for the calculation and analysis of these charm mesons states. For example, $^{3}P_{0}$ model \cite{Sun:PP88,GodfreyMoats:PP88,TianZhang:PP88}, Chiral quark model \cite{Zhong:PP88}, lattice QCD model \cite{Moir:PP88,Lewis:PP88,MoirRTh:PP88}, other models \cite{EbertFaustov:PP888,AsgharM:PP888,GodfreySw:PP888,PanWang:PP888,MasudSwa:PP888}, etc. In addition, the high energy $B$, $B_{s}$ mesons \cite{ZengVan:PP888,LahdeN:PP888,PierroE:PP888,Chen:PP88,Chen:PP888} have been extensively studied, there are different interpretations for the bottom meson states, such as $B_{J}(5840)$, $B^{\ast}_{J}(5970)$ \cite{Zylae:PP888,Aaije:PP888,AsgharMasud:PP888,YuWang:PP888,ChenLuo:PP888,FengHao:PP888}. So far, these quantum states are still controversial and need to be further confirmed by experiments.

In this paper, The purpose of this work is to calculate the masses of $1S$-and $2S$-wave $D/D_{s}$ and $B/B_{s}$ mesons states from the linear Regge trajectory (spin independent mass) formula and the spin dependent potential (spin dependent mass) with the corresponding spin-parity quantum numbers $J^{P}=0^{-}$ or $1^{-}$, respectively. In Table I, we list the heavy quark and singly light quark masses of charm and bottom mesons, and the string tensions $\alpha(Q\bar q)$.

\renewcommand\tabcolsep{0.47cm}
\renewcommand{\arraystretch}{1.0}
\begin{table*}[!htbp]
\caption{The effective masses(GeV) of quarks determined by Regge trajectory and the relativistic quark model are compared, with $\alpha$ in $GeV^{-2}$. }\label{em}
\begin{tabular}{cccccccccc}
\hline\hline
{\small Parameters} & ${\small M}_{c}$ & ${\small M}_{b}$ & ${\small m}_{n}$
& ${\small m}_{s}$ & ${\small \alpha}${\small (}${\small c\bar{n}}${\small )} & $%
{\small \alpha}${\small (}${\small c\bar{s}}${\small )} & ${\small \alpha(b\bar{n})}$
& ${\small \alpha(b\bar{s})}$ \\ \hline
{\small Ref.\cite{Jia:PP88}} & ${\small 1.44}$ & ${\small 4.48}$ & ${\small 0.23}$ & $%
{\small 0.328}$ & ${\small 0.223}$ & ${\small 0.249}$ & ${\small 0.275}$ & $%
{\small 0.313}$ \\
{\small EFG \cite{EFG:C10}} & ${\small 1.55}$ & ${\small 4.88}$ & ${\small 0.33}$ & $%
{\small 0.5}$ & ${\small 0.64}${\small /}${\small 0.58}$ & ${\small 0.68/0.64%
}$ & ${\small 1.25/1.21}$ & ${\small 1.28/1.23}$ \\
\hline\hline
\end{tabular}
\end{table*}

 This paper is organized as follows. We analyze the Regge trajectory formula to give $S$-wave spin-average masses in Sec. II. In Sec. III, we review about the spin-dependent Hamiltonian. In Sec. IV, we talk about the scaling relations. In Sec. V, we mainly  employ the Breit-Fermi spin interaction to calculate
their spin coupling parameters and wave functions . We present conclusions in Sec. VI.

\section{The Regge trajectory and the spin average masses}\label{Sec.II}

Considering the color interaction between quarks and quarks in the heavy-light mesons system, in order to estimate the masses splitting in orbitally excited charm and bottom mesons , we use Regge-like masses relation \cite{ChenDLM:PP88} to comprehensively analyze the whole singly heavy system of mesons,
\begin{equation}
(\bar M_{L}-M_{Q})^{2}=\alpha\pi L+a_{0} , \label{pp1}
\end{equation}
where, $\bar M_{L}$, $M_{Q} $ are the spin-average mass and the heavy quark mass of the singly heavy meson, respectively. $L$ is the orbital angular momentum of the mesons system($L=0, 1, 2, \cdot\cdot\cdot$). $\alpha$ is the QCD string tension coefficient between heavy quark and light antiquark. The intercept factor $a_{0}$ depends on
the light antiquark mass $m_{\bar q}$ and the non-relativistic kinematic energy $P^{2}_{Q}/M_{Q}$ of the heavy quark,
\begin{equation}
a_{0}=(m_{\bar q}+\frac{P^{2}_{Q}}{M_{Q}})^{2} , \label{pp2}
\end{equation}
note that non-relativistic kinematic $3-$momentum of heavy quark in Eq. (\ref{pp2}) has been associated with both $M_{Q}$ and $v_{Q}$,
\begin{eqnarray}
P_{Q}\equiv M_{Q}v_{Q} \ , \  \  \ \  \  \  \  \ v_{Q}=(1-\frac{m_{bareQ}^{2}}{M_{Q}^{2}})^{\frac{1}{2}}, \label{pp3}
\end{eqnarray}
here, $v_{Q}$ is the velocity of the heavy quark, and the $3-$momentum $P_{Q}$ is conserved in the heavy quark limit of $M_{Q}\rightarrow \infty$. Using Eqs. (\ref{pp1}),(\ref{pp2}) and (\ref{pp3}),one can obtain the spin-averaged masses \cite{Jia:PP88,JiaPan:PP888} by
{\small{\begin{equation}
\begin{aligned}
\bar M_{L}&=&M_{Q}+\sqrt{\alpha\pi L+\left( m_{\bar q}+M_{Q}\left( 1-\frac{m_{bareQ}^{2}}{M_{Q}^{2}}\right) \right) ^{2}}\text{,} \label{pp421}
\end{aligned}
\end{equation}}}
where, $m_{bareQ}$ and $m_{\bar q}$ are the heavy quark bare mass of singly heavy meson and the singly light antiquark mass, respectively. The selection of these parameters is listed in Table I.

For the singly heavy mesons, it is interesting to note that in \cite{JiaDH:PP888} this Regge trajectory is proposed in consideration that the trajectory slope ratio between the radial and angular-momentum Regge trajectories to be $\pi/2:1$. Then,$\pi\alpha L$ in Eq. (\ref{pp421}) is replaced by $\pi\alpha(L+\frac{\pi}{2}n)$,
{\small{\begin{equation}
\begin{aligned}
\bar M_{L}&=&M_{Q}+\sqrt{\alpha\pi (L+\frac{\pi}{2}n)+\left( m_{\bar q}+M_{Q}\left( 1-\frac{m_{bareQ}^{2}}{M_{Q}^{2}}\right) \right) ^{2}}\text{,} \label{pp4}
\end{aligned}
\end{equation}}}
where, $n$ is the radial quantum number. By analogy of the heavy meson system, we estimate the S-wave the spin-average mass of the $Q(\bar q)$ state using Eq. (\ref{pp4}). For instance, we get the $1S$-wave spin-average mass $(L=0,n=0)$ in Eq. (\ref{pp4}) is \ \ $\bar M_{L}=M_{Q}+\left( m_{\bar q}+M_{Q}\left( 1-\frac{m_{bareQ}^{2}}{M_{Q}^{2}}\right) \right)$, and the $2S$-wave spin-average mass $(L=0,n=1)$ is \ \ $\bar M_{L}=M_{Q}+\sqrt{\alpha\pi(\pi/2)+\left( m_{\bar q}+M_{Q}\left( 1-\frac{m_{bareQ}^{2}}{M_{Q}^{2}}\right) \right) ^{2}}$.  Given trajectory parameters in Table I, one can predict Regge trajectories of the charm excited mesons $D/D_{s}$ in FIG. 1-2 and the bottom excited mesons $B/B_{s}$ in FIG. 3-4 with the radial quantum number $n = 0$, $1$, $2$, $3$ and $4$ .

\begin{figure*}
\centering
\begin{minipage}[t]{0.49\textwidth}
   % \includegraphics[width=5.6cm]{nucleveropro}
\includegraphics[width=6cm]{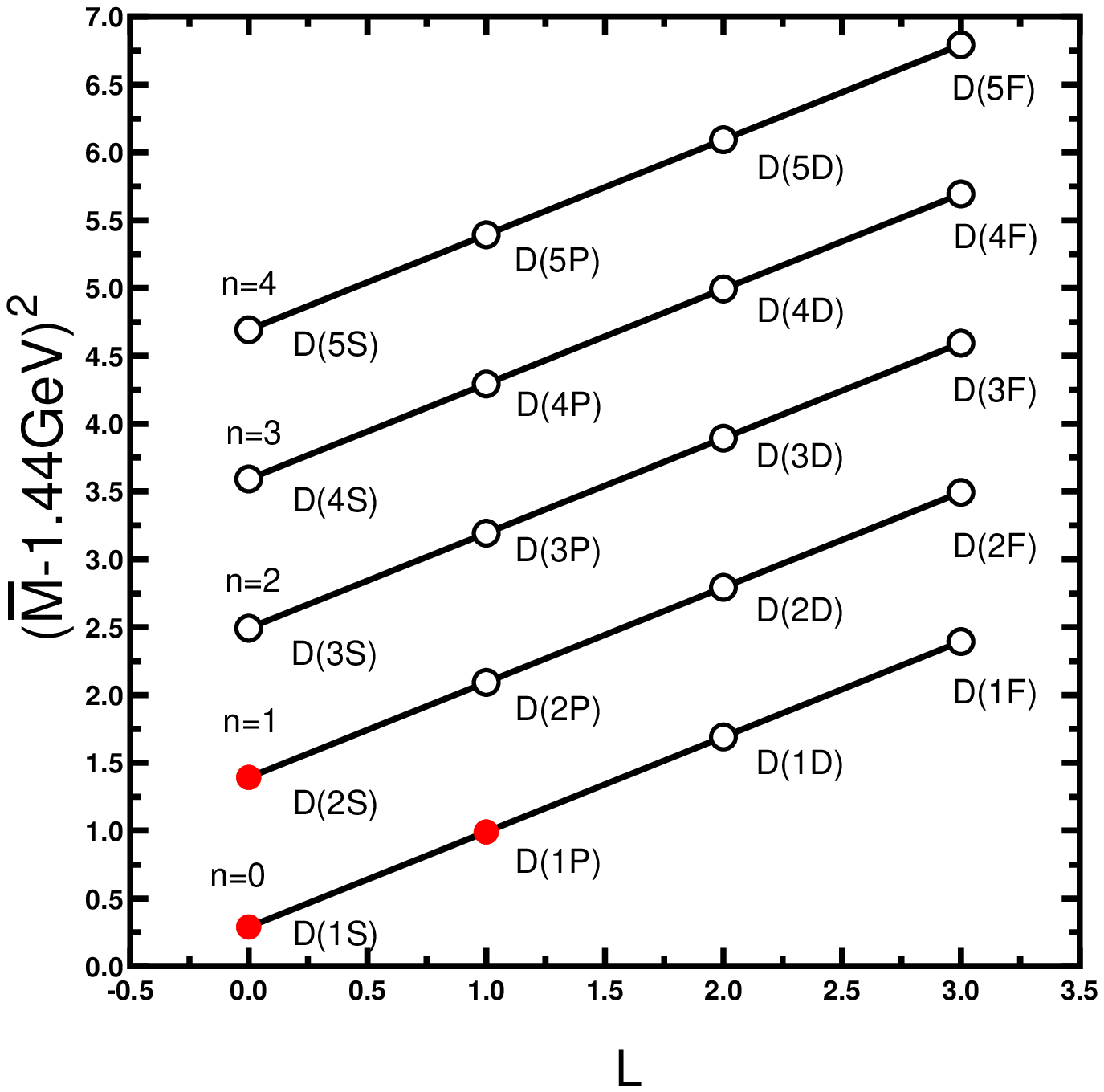}
\caption{$D$ meson spin-average mass}
\includegraphics[width=6cm]{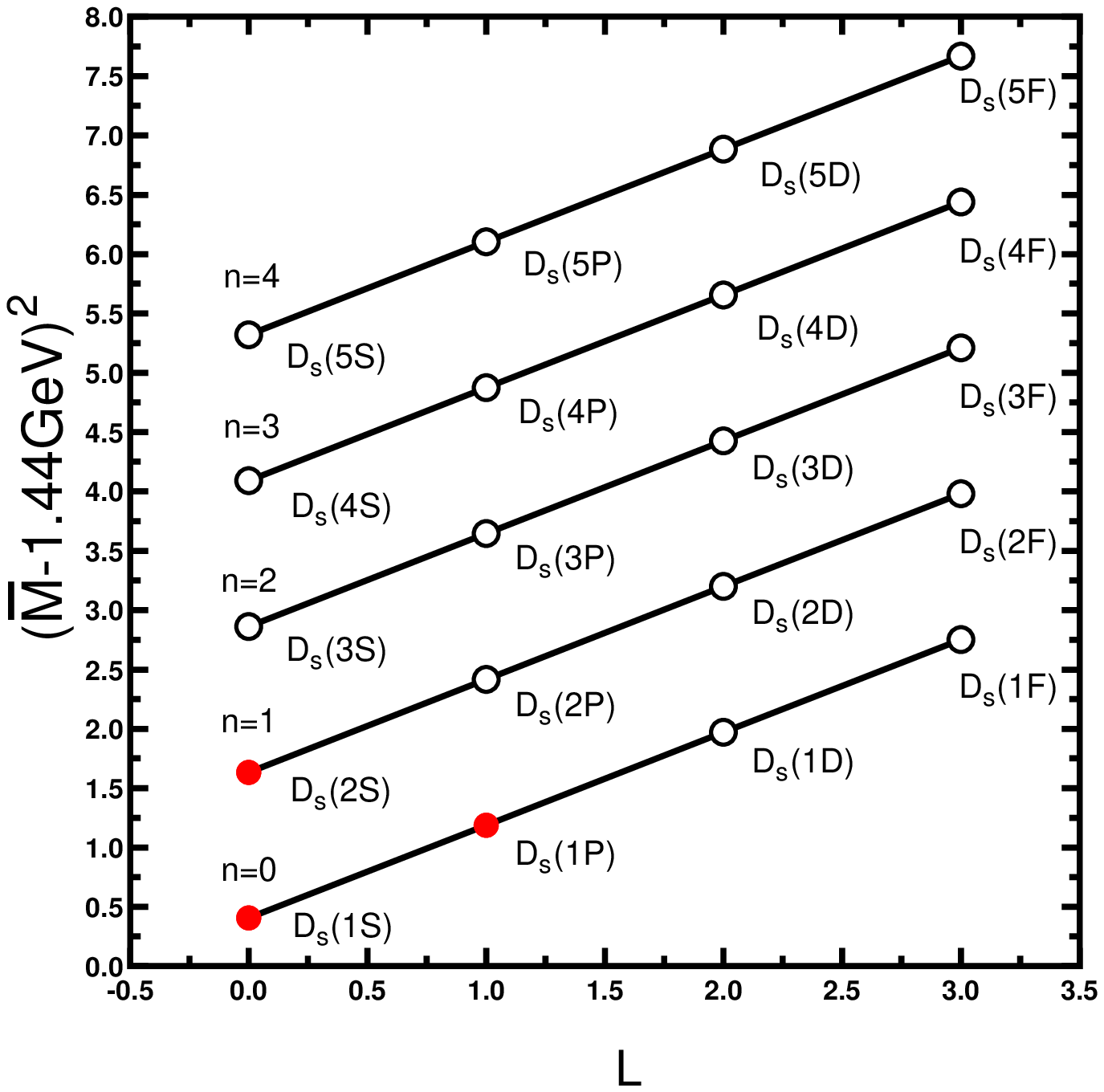}
\caption{$D_{s}$ meson spin-average mass}
\end{minipage}
\begin{minipage}[t]{0.49\textwidth}
\includegraphics[width=6cm]{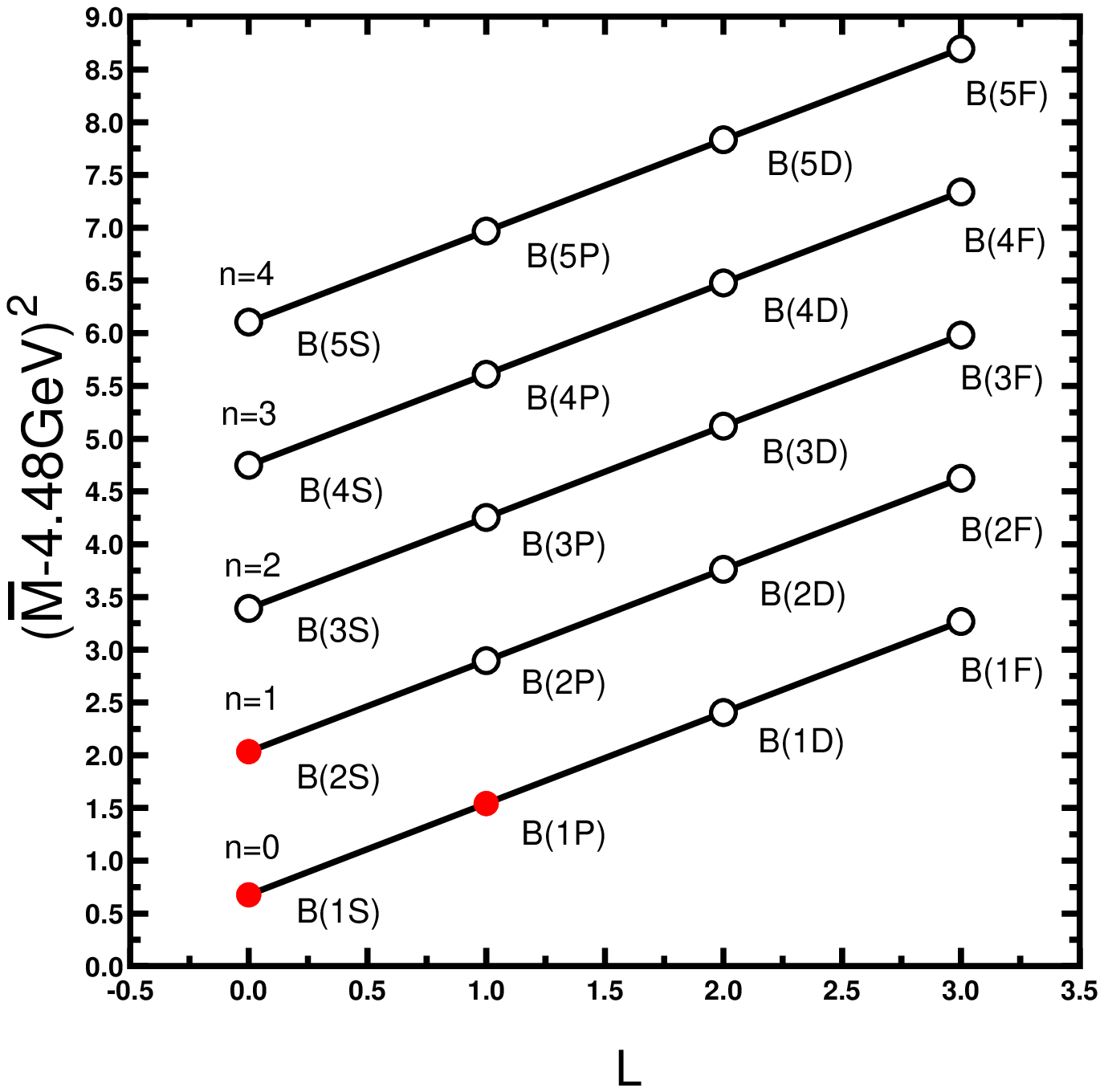}
\caption{$B$ meson spin-average mass}
\includegraphics[width=6cm]{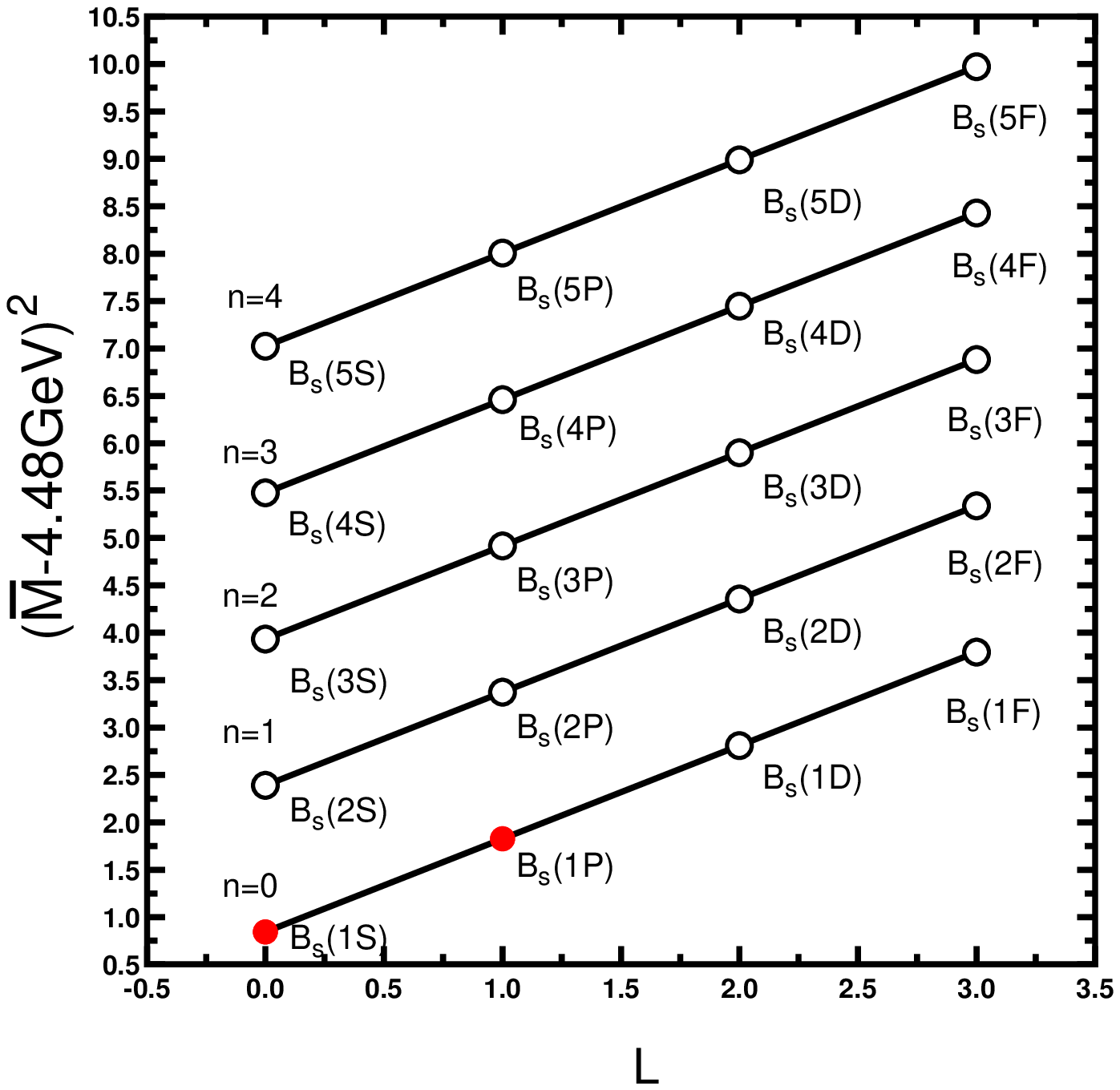}
\caption{$B_{s}$ meson spin-average mass}
\end{minipage}
\end{figure*}

Regge trajectories of the heavy-light hadron systems relating the shifted spin-averaged mass squared
to the orbital angular momentum $L$ of the systems, with the parameters in Table I corresponding to Eq. (\ref{pp4}). The red solid circles correspond to the observed (mean) masses, and the empty circles indicate the predicted value in FIG. $1-4$.

\section{The spin-dependent potential}\label{Sec.III}

Due to the spin-spin interaction between singly light antiquark and heavy quark of singly heavy mesons, in order to estimate the mass splitting, we consider the spin-dependent Hamiltonian $H^{SD}$ \cite{EbertFGS:PP888,KarlinerRP:PP888} ,
\begin{equation}
H^{SD}=a_{1}\mathbf{L}\cdot \mathbf{S}_{\bar q}+a_{2}\mathbf{L}\cdot \mathbf{S}%
_{Q}+bS_{12}+c\mathbf{S}_{\bar q}\cdot \mathbf{S}_{Q},  \label{PP5}
\end{equation}%
\begin{equation*}
S_{12}=3(\mathbf{S}_{\bar q}\cdot \mathbf{\hat{r}})(\mathbf{S}_{Q}\cdot \mathbf{\hat{r}})/r^{2}-%
\mathbf{S}_{\bar q}\cdot \mathbf{S}_{Q}\mathbf{,}
\end{equation*}%
where, the first two terms are spin-orbit interactions, the third is the
tensor energy, and the last is the contact interaction between the heavy quark
spin $\mathbf{S}_{Q}$ and the antiquark spin $\mathbf{S}_{\bar q}$. Here, $a_{1}$, $a_{2}$, $b$, $c$ are spin coupling parameters.

For the singly heavy mesons, the singly light antiquark spin and the heavy quark
spin are ${S}_{\bar{q}}=\frac{1}{2}$, ${S}_{Q}=\frac{1}{2}$, respectively. Therefore, there are two kinds of the total spin $S$, one is 0 and the other is 1. In the scheme of $LS$ coupling,  note that the total angular momentum $J=S+L$. Coupling of $L = 0$ with the spin $S = 0$ gives states with the total angular momentum $J = 0$, while coupling with $S = 1$
leads to states the angular momentum $J =1$.

We consider the $S$-wave($L = 0$) states in $Q\bar q$ mesons case. Then, the first three terms of in Eq. (\ref{PP5}) are eliminated, only the last term survives,
\begin{equation}
H^{SD}=c\mathbf{S}_{\bar q}\cdot \mathbf{S}_{Q}.  \label{PP6}
\end{equation}
It is very convenient to analyze spin-spin interaction into the non-trivial terms for the mass splitting, the eigenvalues (two diagonal elements) of $<\mathbf{S}_{\bar q}\cdot \mathbf{S}_{Q}>$ can be obtained,
\begin{equation}
<\mathbf{S}_{\bar q}\cdot \mathbf{S}_{Q}>=[S(S+1)-S_{Q}(S_{Q}+1)-S_{\bar q}(S_{\bar q}+1)]/2,  \label{PP7}
\end{equation}
\begin{equation}
<\mathbf{S}_{\bar q}\cdot \mathbf{S}_{Q}>=\left[
\begin{array}{cc}
-\frac{3}{4} & 0 \\
0 & \frac{1}{4} \label{pp8}
\end{array}
\right] ,
\end{equation}%
combining with Eqs. (\ref{pp4}) and (\ref{pp8}), the $S$-wave masses are,
\begin{equation}
M(S)=\bar M_{L}+c\left[
\begin{array}{cc}
-\frac{3}{4} & 0 \\
0 & \frac{1}{4} \label{pp9}
\end{array}
\right] .
\end{equation}%

Using the above Eq.(\ref{pp9}), we can calculate the mass spectrum of the charm and the bottom mesons states $(D/D_{s}, B/B_{s})$. There is much discussion about the hyperfine splitting terms, and here we mainly determine the parameter values and then derive the effective mass of the mesons.

\section{The hyperfine splitting parameters}\label{Sec.IV}

\begin{figure*}
\centering
\begin{minipage}[t]{0.49\textwidth}
   % \includegraphics[width=5.6cm]{nucleveropro}
\includegraphics[width=8cm]{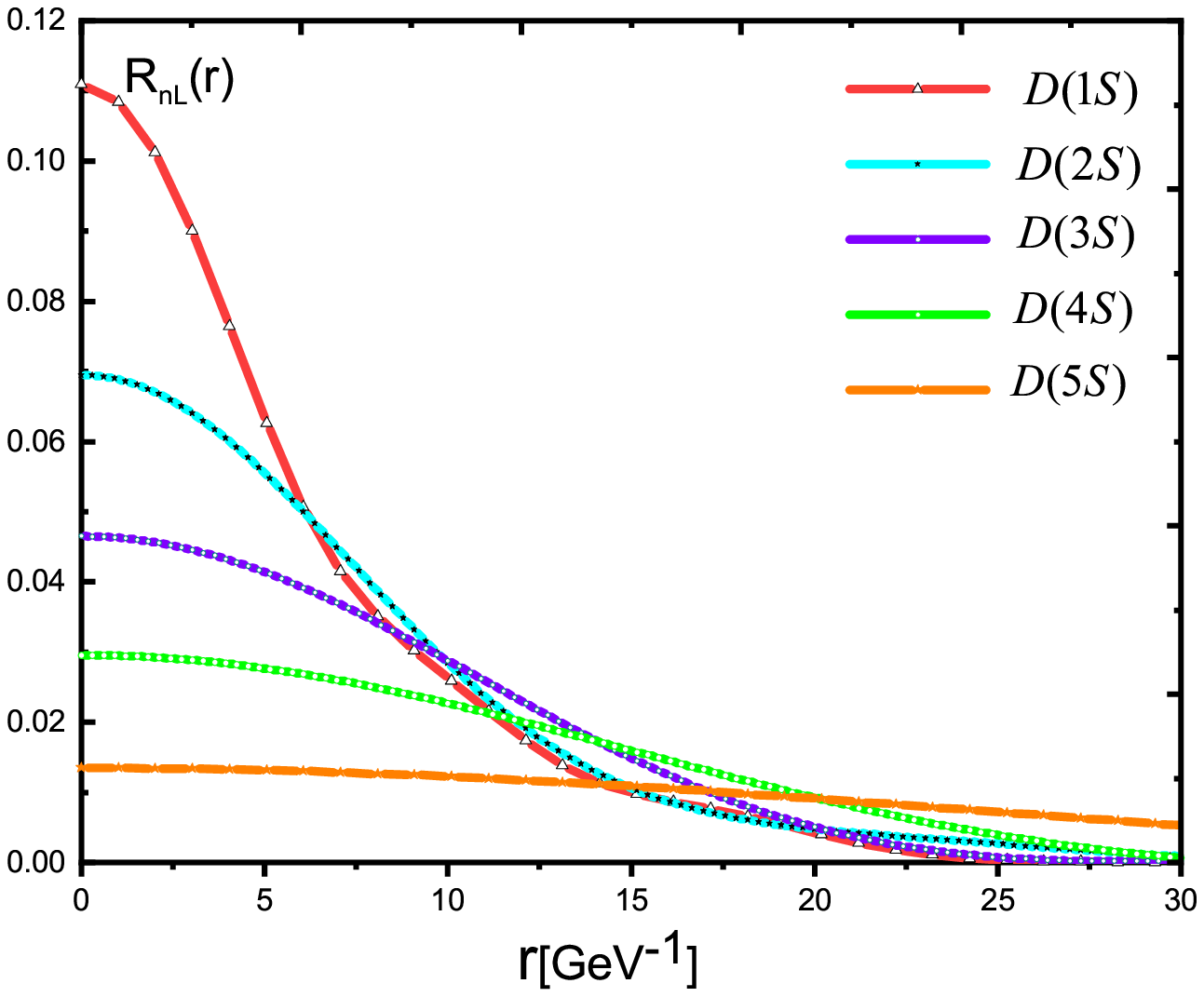}
\caption{$D$ meson radial wavefunction}
\includegraphics[width=8cm]{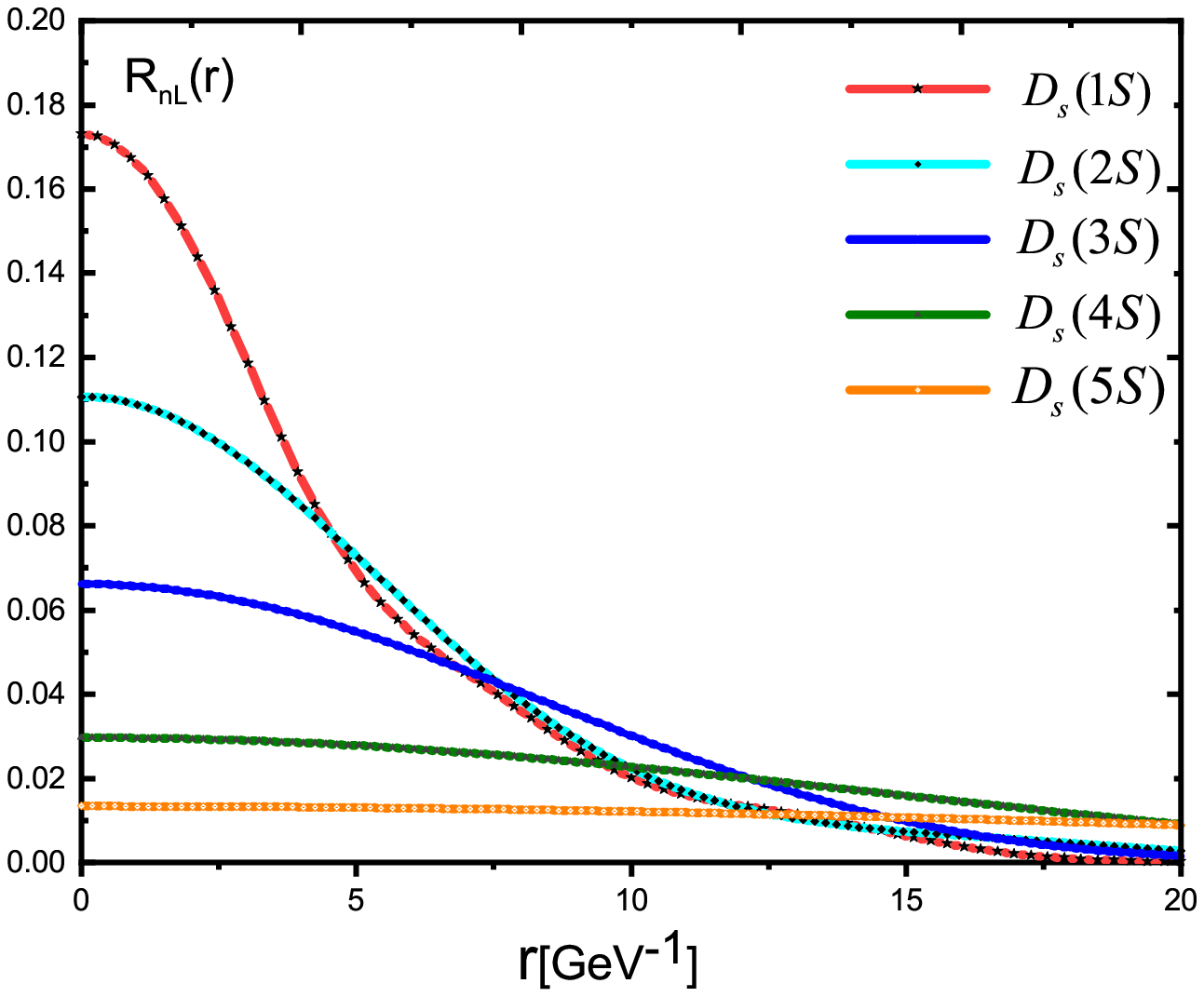}
\caption{$D_{s}$ meson radial wavefunction}
\end{minipage}
\begin{minipage}[t]{0.49\textwidth}
\includegraphics[width=8cm]{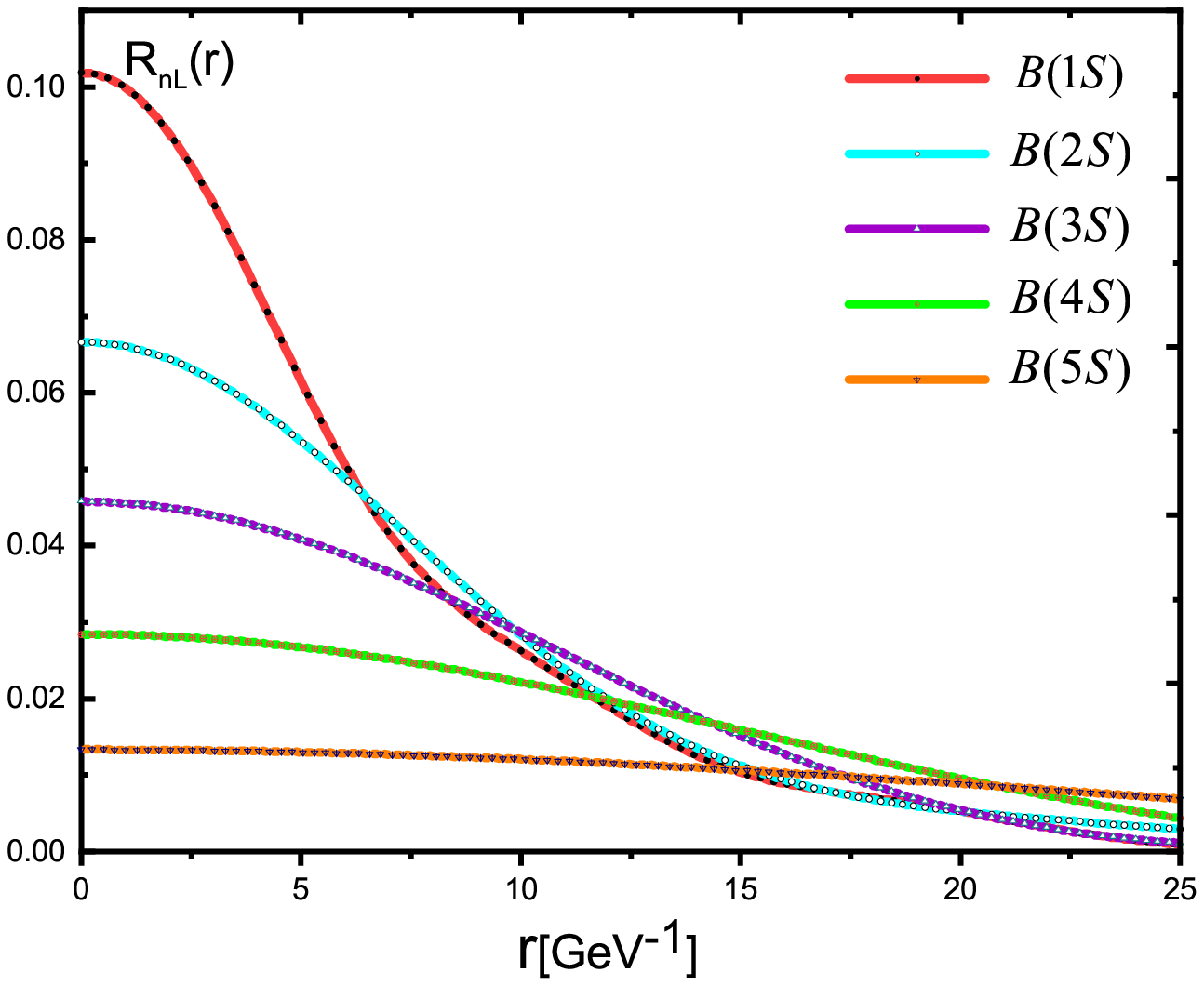}
\caption{$B$ meson radial wavefunction}
\includegraphics[width=8cm]{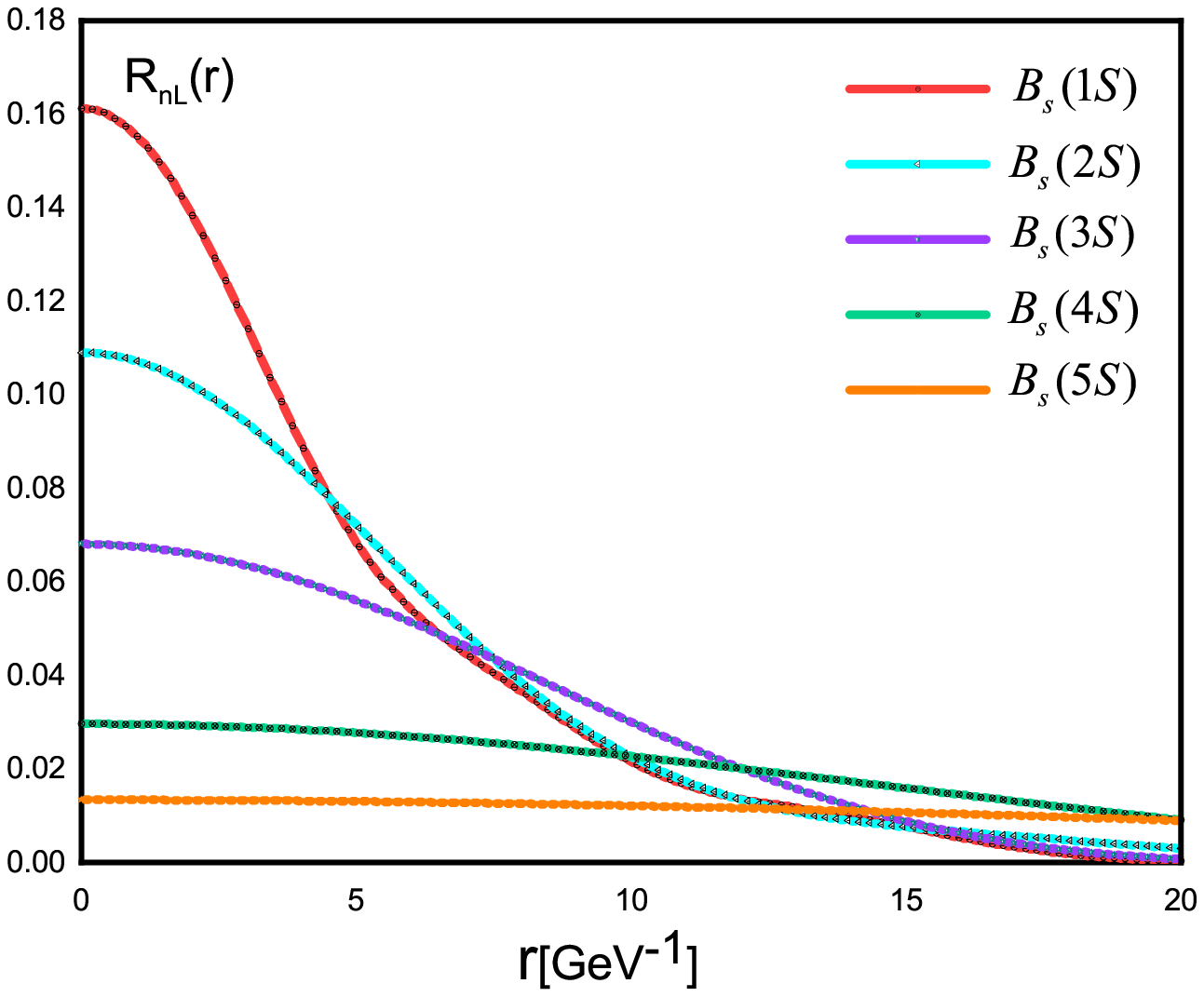}
\caption{$B_{s}$ meson radial wavefunction}
\end{minipage}
\end{figure*}
It is possible to apply the charm and bottom mesons masses of $M(D,1S)=1869.59 MeV$, $M(D^{\ast},1S)=2010.26 MeV$ and $M(D_{s},1S)=1968.3 MeV$, $M(D_{s}^{\ast},1S)=2112.2 MeV$ \cite{Tanabashi:PP88}. A linear relationship between them is their spin-weighted sum: $ \mathop{\sum}\limits_{J}(2J+1)\Delta \bar M=0$,
\begin{eqnarray}
\bar M_{D, 1S}&=&\frac{1869.59 MeV+3\times2010.26 MeV}{4} \notag \\
&=&1975.09 MeV,  \label{pp10}
\end{eqnarray}
\begin{eqnarray}
\bar M_{D_{s}, 1S}&=&\frac{1968.3 MeV+3\times2112.2 MeV}{4}\notag\\
&=&2076.23 MeV, \label{pp11}
\end{eqnarray}
the spin-averaged masses calculated by Eq. (\ref{pp4}) are nearly equal to the results of the experimental value, namely, Eq. (\ref{pp10}) and  Eq. (\ref{pp11}), respectively. In addition, the mass splitting are
\begin{eqnarray}
c(D, 1S)&=&M(1 ^{3}S_{1},1^{-})-M(1 ^{1}S_{0},0^{-}) \notag \\
&=&140.6MeV  \label{pp12},
\end{eqnarray}
\begin{eqnarray}
c(D_{s}, 1S)&=&M(1 ^{3}S_{1},1^{-})-M(1 ^{1}S_{0},0^{-})\notag\\
&=&143.9MeV  \label{pp13}.
\end{eqnarray}

From Eqs. (\ref{pp12}) and (\ref{pp13}), it is assumed that $c(D, 1S)\approx c(D_{s}, 1S)$ is about $140MeV$ in the charm meson of the $1S$-wave. However,  one can use a similar approach in the $1S$-wave bottom mesons $(B/B_{s})$, corresponding to the scaling relations from the charm meson $(c\bar n)$ and then to the bottom meson $(b\bar s)$, as well as the following rough estimate of the parameter $c$,
\begin{equation}
c(b\bar s)\approx c(b\bar n)  \label{pp14}.
\end{equation}

To estimate the mass of the bottom flavor meson with vector singly light quarks (spin $S_{\bar q} = 1 / 2$), we use the spin coupling parameters obtained from experimental measurements $D$ and $D_{s}$ mesons values, which should theoretically have an approximate mapping with the following scaling relationship,
\begin{equation}
c(b\bar q)=\frac{M_{c}}{M_{b}}c(c\bar q),  \label{pp15}
\end{equation}
with the heavy quarks relation $M_{c}/M_{b}$ for the parameter of $c(B,1S), c(B_{s},1S)$ are about 45 MeV. We will show the results of our calculations, as compared with the experimental values, with reasonable values of the parameters achieved.

\section{Spin-spin couplings in relativized quark model}\label{Sec.V}

\renewcommand\tabcolsep{0.01cm}
\renewcommand{\arraystretch}{1.0}
\begin{table*}[!htbp]
\caption{Model Parameters(in GeV).   \label{Table 8}}
\[\begin{array}{ccccccccc}
\hline\hline
{\small State}   & \alpha_{s} & \sigma   & c_{1}   &c_{2}&   c_{3}&   c_{4}&   c_{5} \\
\hline
\small D    &0.564     & 0.230       & 0.143  &0.089  &0.049  &0.022 &0.005  \\
\small D_{s} &0.537   & 0.260     & 0.144  & 0.100  & 0.048 &0.012 &0.003 \\
\small B    &0.532     & 0.216      & 0.041    & 0.027  &0.016 &0.007 &0.002 \\
\small B_{s} &0.526    &0.200      & 0.048    &0.038   & 0.022 &0.006 &0.001 \\
\hline\hline
\end{array}\]
\end{table*}

\renewcommand\tabcolsep{0.15cm}
\renewcommand{\arraystretch}{1.0}
\begin{table*}[!htbp]
\caption{The S-wave mass spectrum (MeV) of the charmed and bottomed mesons are given and compared with the different quark models.}\label{dm}
\begin{tabular}{ccccccccccc}

\hline\hline
{\small State }$J^{P}$ & {\small Meson} & {\small Mass} &{Ours}& GI \cite{GodfreySw:PP888,GodfreyMo:D85}&{\small EFG \cite{EFG:C10}}& {\small Meson} & {\small Mass} &{Ours}& GI \cite{GodfreySw:PP888,GodfreyMo:D85}&{\small EFG \cite{EFG:C10}} \\
\hline
$%
\begin{array}{rr}
{\small 1}^{1}{\small S}_{0} & {\small 0}^{-} \\
{\small 1}^{3}{\small S}_{1} & {\small 1}^{-}%
\end{array}%
$ & $%
\begin{array}{r}
{\small D}^{\pm } \\
{\small D}^{\ast }%
\end{array}%
$ & $%
\begin{array}{r}
{\small 1869.59} \\
{\small 2010.26}%
\end{array}%
$ & $%
\begin{array}{r}
{\small 1870.86} \\
{\small 2014.13}%
\end{array}%
$ & $%
\begin{array}{r}
{\small 1877} \\
{\small 2041}%
\end{array}%
$ & $%
\begin{array}{r}
{\small 1871} \\
{\small 2010}%
\end{array}%
$ & $%
\begin{array}{r}
{\small D}_{s} \\
{\small D}_{s}^{\ast }%
\end{array}%
$ & $%
\begin{array}{r}
{\small 1968.3} \\
{\small 2112.2}%
\end{array}%
$ & $
\begin{array}{r}
{\small 1968.18} \\
{\small 2112.35}%
\end{array}%
$ & $%
\begin{array}{r}
{\small  1979} \\
{\small 2129}%
\end{array}%
$ & $%
\begin{array}{r}
{\small 1969} \\
{\small 2111}%
\end{array}%
$ \\ $%

\begin{array}{rr}
{\small 2}^{1}{\small S}_{0} & {\small 0}^{-} \\
{\small 2}^{3}{\small S}_{1} & {\small 1}^{-}%
\end{array}%
$ & $%
\begin{array}{r}
{\small D}_{0} \\
{\small D}^{\ast}%
\end{array}%
$ & $%
\begin{array}{r}
{\small 2549\pm16} \\
{\small 2637\pm2}%
\end{array}%
$ & $%
\begin{array}{r}
{\small 2551.58} \\
{\small 2640.86}%
\end{array}%
$ & $%
\begin{array}{r}
{\small 2581} \\
{\small 2643}%
\end{array}%
$ & $%
\begin{array}{r}
{\small 2581} \\
{\small 2632}%
\end{array}%
$ & $%
\begin{array}{r}
{\small } \\
{\small D}^{\ast}_{s1}%
\end{array}%
$ & $%
\begin{array}{r}
{\small } \\
{\small 2708.3}%
\end{array}%
$ & $
\begin{array}{r}
{\small 2642.42} \\
{\small 2742.72}%
\end{array}%
$ & $%
\begin{array}{r}
{\small 2673} \\
{\small  2732}%
\end{array}%
$ & $%
\begin{array}{r}
{\small 2688} \\
{\small 2731}%
\end{array}%
$ \\ $%

\begin{array}{rr}
{\small 3}^{1}{\small S}_{0} & {\small 0}^{-} \\
{\small 3}^{3}{\small S}_{1} & {\small 1}^{-}%
\end{array}%
$ & $%
\begin{array}{r}
{\small} \\
{\small}%
\end{array}%
$ & $%
\begin{array}{r}
{\small } \\
{\small }%
\end{array}%
$ & $%
\begin{array}{r}
{\small 2981.16} \\
{\small 3029.66}%
\end{array}%
$ & $%
\begin{array}{r}
{\small 3068} \\
{\small 3110}%
\end{array}%
$ & $%
\begin{array}{r}
{\small 3062} \\
{\small 3096}%
\end{array}%
$ & $%
\begin{array}{r}
{\small } \\
{\small }%
\end{array}%
$ & $%
\begin{array}{r}
{\small } \\
{\small }%
\end{array}%
$ & $
\begin{array}{r}
{\small 3094.93} \\
{\small 3143.35}%
\end{array}%
$ & $%
\begin{array}{r}
{\small  3154} \\
{\small  3193}%
\end{array}%
$ & $%
\begin{array}{r}
{\small 3219} \\
{\small 3242}%
\end{array}%
$ \\ $%

\begin{array}{rr}
{\small 4}^{1}{\small S}_{0} & {\small 0}^{-} \\
{\small 4}^{3}{\small S}_{1} & {\small 1}^{-}%
\end{array}%
$ & $%
\begin{array}{r}
{\small } \\
{\small }%
\end{array}%
$ & $%
\begin{array}{r}
{\small } \\
{\small }%
\end{array}%
$ & $%
\begin{array}{r}
{\small 3318.03} \\
{\small 3339.75}%
\end{array}%
$ & $%
\begin{array}{r}
{\small 3468} \\
{\small 3497}%
\end{array}%
$ & $%
\begin{array}{r}
{\small 3452} \\
{\small 3482}%
\end{array}%
$ & $%
\begin{array}{r}
{\small } \\
{\small }%
\end{array}%
$ & $%
\begin{array}{r}
{\small } \\
{\small }%
\end{array}%
$ & $
\begin{array}{r}
{\small 3453.16} \\
{\small 3464.92}%
\end{array}%
$ & $%
\begin{array}{r}
{\small  3547} \\
{\small  3575}%
\end{array}%
$ & $%
\begin{array}{r}
{\small 3652} \\
{\small 3669}%
\end{array}%
$ \\ $%

\begin{array}{rr}
{\small 5}^{1}{\small S}_{0} & {\small 0}^{-} \\
{\small 5}^{3}{\small S}_{1} & {\small 1}^{-}%
\end{array}%
$ & $%
\begin{array}{r}
{\small } \\
{\small }%
\end{array}%
$ & $%
\begin{array}{r}
{\small } \\
{\small }%
\end{array}%
$ & $%
\begin{array}{r}
{\small 3601.57} \\
{\small 3606.49}%
\end{array}%
$ & $%
\begin{array}{r}
{\small 3814} \\
{\small 3837}%
\end{array}%
$ & $%
\begin{array}{r}
{\small 3793} \\
{\small 3822}%
\end{array}%
$ & $%
\begin{array}{r}
{\small } \\
{\small}%
\end{array}%
$ & $%
\begin{array}{r}
{\small } \\
{\small }%
\end{array}%
$ & $
\begin{array}{r}
{\small 3743.86} \\
{\small 3746.42}%
\end{array}%
$ & $%
\begin{array}{r}
{\small   3894} \\
{\small   3912}%
\end{array}%
$ & $%
\begin{array}{r}
{\small 4033} \\
{\small 4048}%
\end{array}%
$ \\
\hline

$%
\begin{array}{rr}
{\small 1}^{1}{\small S}_{0} & {\small 0}^{-} \\
{\small 1}^{3}{\small S}_{1} & {\small 1}^{-}%
\end{array}%
$ & $%
\begin{array}{r}
B^{0} \\
B^{\ast }%
\end{array}%
$ & $%
\begin{array}{r}
{\small 5279.6} \\
{\small 5324.7}%
\end{array}%
$ & $
\begin{array}{r}
{\small 5279.17} \\
{\small 5320.27}%
\end{array}%
$ & $%
\begin{array}{r}
{\small  5312} \\
{\small  5371}%
\end{array}%
$ & $%
\begin{array}{r}
{\small 5280} \\
{\small 5326}%
\end{array}%
$ & $
\begin{array}{r}
{\small B}_{s} \\
{\small B}_{s}^{\ast }%
\end{array}%
$ & $%
\begin{array}{r}
{\small 5366.9} \\
{\small 5414.4}%
\end{array}%
$ & $
\begin{array}{r}
{\small 5366.58} \\
{\small 5414.15}%
\end{array}%
$ & $%
\begin{array}{r}
{\small  5394} \\
{\small  5450}%
\end{array}%
$ & $%
\begin{array}{r}
{\small 5372} \\
{\small 5414}%
\end{array}%
$ \\ $%

\begin{array}{rr}
{\small 2}^{1}{\small S}_{0} & {\small 0}^{-} \\
{\small 2}^{3}{\small S}_{1} & {\small 1}^{-}%
\end{array}%
$ & $%
\begin{array}{r}
B_{J}\\
B_{J}%
\end{array}%
$ & $%
\begin{array}{r}
{\small 5851\pm19} \\
{\small 5964\pm5}%
\end{array}%
$ & $
\begin{array}{r}
{\small 5892.94} \\
{\small 5920.12}%
\end{array}%
$ & $%
\begin{array}{r}
{\small 5904} \\
{\small 5933}%
\end{array}%
$ & $%
\begin{array}{r}
{\small 5890} \\
{\small 5906}%
\end{array}%
$ & $%
\begin{array}{r}
 \\
\end{array}%
$ & $%
\begin{array}{r}
{\small } \\
{\small }%
\end{array}%
$ & $
\begin{array}{r}
{\small 6000.83} \\
{\small 6039.26}%
\end{array}%
$ & $%
\begin{array}{r}
{\small 5984} \\
{\small 6012}%
\end{array}%
$ & $%
\begin{array}{r}
{\small 5976} \\
{\small 5992}%
\end{array}%
$ \\ $%

\begin{array}{rr}
{\small 3}^{1}{\small S}_{0} & {\small 0}^{-} \\
{\small 3}^{3}{\small S}_{1} & {\small 1}^{-}%
\end{array}%
$ & $%
\begin{array}{r}
 \\
\end{array}%
$ & $%
\begin{array}{r}
{\small} \\
{\small}%
\end{array}%
$ & $
\begin{array}{r}
{\small 6316.81} \\
{\small 6332.52}%
\end{array}%
$ & $%
\begin{array}{r}
{\small  6335} \\
{\small 6355}%
\end{array}%
$ & $%
\begin{array}{r}
{\small 6379} \\
{\small 6387}%
\end{array}%
$ & $%
\begin{array}{r}
 \\
\end{array}%
$ & $%
\begin{array}{r}
{\small } \\
{\small }%
\end{array}%
$ & $
\begin{array}{r}
{\small 6450.78} \\
{\small 6473.15}%
\end{array}%
$ & $%
\begin{array}{r}
{\small  6410} \\
{\small  6429}%
\end{array}%
$ & $%
\begin{array}{r}
{\small 6467} \\
{\small 6475}%
\end{array}%

$ \\ $%

\begin{array}{rr}
{\small 4}^{1}{\small S}_{0} & {\small 0}^{-} \\
{\small 4}^{3}{\small S}_{1} & {\small 1}^{-}%
\end{array}%
$ & $%
\begin{array}{r}
 \\
\end{array}%
$ & $%
\begin{array}{r}
{\small } \\
{\small }%
\end{array}%
$ & $
\begin{array}{r}
{\small 6661.00} \\
{\small 6667.82}%
\end{array}%
$ & $%
\begin{array}{r}
{\small 6689} \\
{\small 6703}%
\end{array}%
$ & $%
\begin{array}{r}
{\small 6781} \\
{\small 6786}%
\end{array}%
$ & $%
\begin{array}{r}
 \\
\end{array}%
$ & $%
\begin{array}{r}
{\small } \\
{\small }%
\end{array}%
$ & $
\begin{array}{r}
{\small 6820.55} \\
{\small 6826.35}%
\end{array}%
$ & $%
\begin{array}{r}
{\small 6759} \\
{\small 6773}%
\end{array}%
$ & $%
\begin{array}{r}
{\small 6874} \\
{\small 6879}%
\end{array}%
$ \\ $%

\begin{array}{rr}
{\small 5}^{1}{\small S}_{0} & {\small 0}^{-} \\
{\small 5}^{3}{\small S}_{1} & {\small 1}^{-}%
\end{array}%
$ & $%
\begin{array}{r}
 \\
\end{array}%
$ & $%
\begin{array}{r}
{\small} \\
{\small}%
\end{array}%
$ & $
\begin{array}{r}
{\small 6956.72} \\
{\small 6958.36}%
\end{array}%
$ & $%
\begin{array}{r}
{\small  6997} \\
{\small 7008}%
\end{array}%
$ & $%
\begin{array}{r}
{\small 7129} \\
{\small 7133}%
\end{array}%
$ & $%
\begin{array}{r}
 \\
\end{array}%
$ & $%
\begin{array}{r}
{\small } \\
{\small }%
\end{array}%
$ & $
\begin{array}{r}
{\small 7133.48} \\
{\small 7134.81}%
\end{array}%
$ & $%
\begin{array}{r}
{\small   7063} \\
{\small   7076}%
\end{array}%
$ & $%
\begin{array}{r}
{\small 7231} \\
{\small 7235}%
\end{array}%
$ \\
\hline\hline
\end{tabular}
\end{table*}

In this section, we mainly analyze the mass splitting of the singly heavy meson
in a relativistic quark model, in which the Breit-Fermi spin interaction is used to calculate
the spin coupling parameter $c$ and compare it with the experimentally matched ones.

For this, we consider the quasi-static potential of the Brett-Fermi spin interaction \cite{Godfrey:PP88,Yndurain83} is
\begin{equation}
\begin{array}{r}
V^{\text{quasi-static}}_{L}=V(r)+S(r)+\left( \frac{V^{\prime }-S^{\prime }}{r}\right)
\mathbf{L}\cdot \left( \frac{\mathbf{S}_{1}}{2m_{\bar q}^{2}}+\frac{\mathbf{S}_{2}%
}{2M_{Q}^{2}}\right)  \\
+\left( \frac{V^{\prime }}{r}\right) \mathbf{L}\cdot \left( \frac{\mathbf{S}%
_{1}+\mathbf{S}_{2}}{m_{\bar q}M_{Q}}\right) +\frac{1}{3m_{\bar q}M_{Q}}\left( \frac{%
V^{\prime }}{r}-V^{\prime \prime }\right) S_{12} \\
+\frac{2}{3m_{\bar q}M_{Q}}\left( \mathbf{\nabla }^{2}V(r)\right) \mathbf{S}%
_{1}\cdot \mathbf{S}_{2},%
\end{array}
\label{Vq11}
\end{equation}%
here, $V$ and $S$ are the respective vector and scalar potentials, and $%
V^{\prime }$, $S^{\prime }$ and $V^{\prime \prime }$ their derivatives. In $S$-wave$(L=0)$, the $\mathbf{L}\cdot \mathbf{S}$ terms will be eliminated, the spin-spin interaction of the last term contains a delta function in configuration space and is treated nonperturbatively in Eq. (\ref{Vq11}),
\begin{eqnarray}
V^{\text{quasi-static}}_{L=0}=- \frac{4\alpha_{s}}{3r}+br+\frac{32\alpha_{s}\pi}{9m_{\overline q}M_{Q}}\tilde{\delta} (r) \mathbf{S}
_{1}\cdot \mathbf{S}_{2}, \label{Vq21}
\end{eqnarray}
where $\alpha_{s}$ is the strong coupling and $b$ is a parameter, one finds
\begin{equation}
c =\frac{32\alpha_{s}\pi}{9m_{\overline q}M_{Q}}\langle\tilde{\delta} (r)\rangle. \label{Vq44}%%
\end{equation}
In Ref. \cite{Barnes:D85}, $\tilde{\delta} (r)= (\frac{\sigma}{\sqrt{\pi}})^{3} e^{-\sigma^{2}r^{2}}$, which is a Gaussian-smeared term with a parameter $\sigma$ and the quantum average $\langle\rangle$ is made over the S-wave wavefunction $R_{nL}(r)$ of the meson($Q\bar q$) system. Then, the S-wave wavefunction  $R_{nL}(r)$ with radial quantum number $n=0, 1, 2, 3, 4$ and $L=0$ of the mesons are shown in Figure $5-8$. The S-wave radial wavefunction  $R_{nL}(r)$ has nodes.

In order to get the masses of the S-wave meson system, there are the second parameters $(\alpha_{s}, \sigma)$ are determined by fitting the mass spectrum, and the values of the parameters $c_{1}$, $c_{2}$, $c_{3}$, $c_{4}$ and $c_{5}$ obtained by fitting with  Eq. (\ref{Vq44}) correspond to $1S$, $2S$, $3S$, $4S$ and $5S$ waves in Table II, respectively. The fitting result of parameters $c$ is consistent with the value calculated by Eqs. (\ref{pp14}) and (\ref{pp15}).

With the help of  the relation Eq. (\ref{Vq44}) for fitting the $S$-wave masses, the results are shown in the Table III. One can find that our results for excited charm mesons $D_{0}(2550)$ and $D^{\ast}(2640)$ are considered as the candidate of the 2S-wave, while our model calculations for the experimentally determined $M(D_{s1}^{\ast}(2700)=2708.3MeV$ meson is slightly $40MeV$ higher. On the other hand, according to the LHCb Collaboration's suggestion \cite{Aaije:PP888},  the $B_{J}(5840)$ and $B_{J}(5970)$ can be ${\small 2}^{1}{\small S}_{0}$ and ${\small 2}^{3}{\small S}_{1}$ states, respectively, the mass splitting is about $110MeV$, which appears relatively large compared to our value of $27MeV$ in Table II.

\section{Summary}

In this paper, one can based on  studying the singly heavy mesons $(D/D_{s})$ and $(B/B_{s})$ of the quark and antiquark bound state system. We mainly combine the linear Regge trajectories and the relativistic quark model, discuss the similarity scaling relationship of the spin coupling parameters, and then revisit the 2S-wave and 3S-wave mass spectrums of the spin parity $J^{P}$ = $0^{-}$ or $1^{-}$, and predict the higher energy $S$-wave states mass spectrum of mesons.

Therefore, comparing the experimental data of heavy-light mesons with the predictions of existing theoretical models will provide favorable evidence for the internal interaction and the internal structure of the mesons.

\end{document}